# Longitudinal radiation force of laser pulses and optics of moving particles

Lubomir M. Kovachev*

*Institute of Electronics, Bulgarian Academy of Sciences, 72 Tzarigradsko choose, 1784 Sofia, Bulgaria*
*lubomirkovach@yahoo.com*

**Abstract:** For a long time transverse and longitudinal optical forces are used for non-contact and noninvasive manipulation of small individual particles. The following question arises: What is the impact of these forces on the ensemble of thousand particles in continuous media? The aim of this work is to find analytical expression of the radiation force and potential densities creating from laser pulse propagating in dielectric media. This allows us to find an effective averaged longitudinal real force at the level of the laser pulse' spot. The obtained force is proportional to initial pulse energy and inversely proportional to its time duration. In the femtosecond region the force becomes strong enough to confine the neutral particles into the pulse envelope and translate them with group velocity. In silica for example, the longitudinal force of a femtosecond pulse is significantly greater than the molecular forces. Thus, the fine ablation in silica with short pulses may be due to this longitudinal force, which breaks down the molecular bonds. Additionally, after confinement into the pulse envelope, the moving particles produce new linear and nonlinear effects. The dipole interaction with the electromagnetic field of the particles captured into the pulse generate at carrier-to-envelope frequency, instead of at the carrying ones. This oscillation is in sub-THz range in gases and in THz in solids. In nonlinear regime instead of third harmonics, the ensemble of moving particles generates at frequency proportional to three times the frequency of the envelope-carrier.

## 1. Introduction

As Ashkin demonstrated [1, 2], it is possible to trap particles by lasers working in cw regime. The analytical expression of the radiation force of one individual particle is obtained in dipole approximation and as it is well known is proportional to the transverse gradient of the square of the electrical field. Recently results of manipulating Rayleigh dielectric particles by optical pulses were obtained [3-8]. The theoretical and experimental results in these investigations show that an additional longitudinal force exists in pulse regime. These studies do not use the fact that the optical response of a short laser pulse is non-stationary. For this reason, the radiation force is presented by the phase velocity of the pulse. Actually, the flow of energy and the pulse envelope propagate with group velocity. This fact must be taken into account in the calculation of the ponder-motor force. The question - what kind of radiation forces applied to ensemble of neutrals from a laser in continuous dielectric media is still open. In this paper we obtain analytical expressions of longitudinal radiation force density and after integration, an effective real force at the level of the pulse width in approximation of first order dispersion. The force density is proportional to the second derivative of pulse time envelope, while the real force at the level of the pulse width is inversely proportional to the pulse time duration. That is why the forces vanish in cw regime, while in the femtosecond region leads to trapping of particles into the pulse envelope. The moving neutral particles admit unexpected linear and nonlinear response, THz generation from dipole interaction at carrier to envelope frequency and new nonlinear evolution of the laser pulses. Our calculations show that even with nJ femtosecond pulse propagating in silica, the longitudinal force is of few order of magnitude greater than the molecular forces. Thus, the fine ablation with short pulses may be a result of this force, which broke the molecular connections at the level of the pulse spot.

## 2. Longitudinal radiation force in media with nonstationary optical response

As shown by Gordon [9], in dipole approximation the radiation force of one Rayleigh particle is Lorentz type force and can be written in the form:

$$\vec{F}_R = \alpha\left[\frac{1}{2}\nabla(\vec{E}^2) + \frac{d}{dt}(\vec{E}\times\vec{H})\right] = \alpha\left[\frac{1}{2}\nabla(\vec{E}^2) + \frac{4\pi}{c}\frac{d}{dt}\vec{S}\right], \quad (1)$$

where $\alpha$ is the atomic polarizability and $\vec{S} = \vec{E}\times\vec{H}$ is the Pointing vector. The first term in the brackets is the well-known gradient force, while the second is associated with the propagation of the pulse energy and is proportional to the Pointing vector. To obtain ponder-motor force density in dielectrics we multiply $\alpha$ by the number of atoms per volume $N$ and additionally use the local field correction:

$$\chi^{(1)} = \frac{N\alpha}{1-\frac{4}{3}\pi N\alpha}, \quad (2)$$

where $\chi^{(1)}$ is the linear susceptibility of the media. The equation for the density force is:

$$\vec{F} = \chi^{(1)}\left[\frac{1}{2}\nabla(\vec{E}^2) + \frac{4\pi}{c}\frac{d}{dt}\vec{S}\right]. \quad (3)$$

One natural way to include the media parameters into the expression of Ponder-Motor (PM) force density (3) is by using the divergence of the Pointing vector:

$$-\frac{c}{4\pi}\nabla\cdot\vec{S} = \frac{1}{4\pi}\left(\vec{E}\cdot\frac{\partial}{\partial t}\vec{D} + \vec{H}\cdot\frac{\partial}{\partial t}\vec{B}\right). \quad (4)$$

In dielectrics with non-stationary linear response function the following relations are fulfilled:

$$\vec{D} = \vec{E} + \vec{P}_{lin}, \quad (5)$$

$$\vec{P}_{lin} = 4\pi\int_0^\infty R^{(1)}(\tau)\vec{E}(t-\tau, r)d\tau, \quad (6)$$

$$\vec{B} = \mu\vec{H}, \quad (7)$$

where $\vec{E}$ and $\vec{H}$ are the electric and magnetic fields, $\vec{D}$ and $\vec{B}$ are the electric and magnetic inductions fields, $\vec{P}_{lin}$ is the linear polarization, $R^{(1)}$ is the linear response function and $\mu = const$ is the magnetic permeability. Let us present the electrical and magnetic fields of a laser pulse by the pulse envelopes and carrying frequency $\omega_0$:

$$\vec{E} = (\vec{A}(x,y,z,t)\exp(i\omega_0 t) + c.c.)/2, \quad (8)$$

$$\vec{H} = \mu\vec{B} = \mu(\vec{C}(x,y,z,t)\exp(i\omega_0 t) - c.c)/2i, \quad (9)$$

where $\vec{A}(x,y,z,t)$ and $\vec{C}(x,y,z,t)$ are the complex amplitudes of the electrical and magnetic fields, correspondingly. After using the Fourier presentation of equations (4) - (9), and developing the product of dielectric constant with frequency $\omega\varepsilon(\omega)$ in Tailor series near $\omega_0$, we obtain the following expression for the divergence of the Pointing vector:

$$\nabla \cdot \vec{S} = -\frac{1}{4v_{gr}}\frac{\partial |\vec{A}(r,t)|^2}{\partial t} - \frac{\mu(\omega_0)}{4c}\frac{\partial |\vec{C}(r,t)|^2}{\partial t}, \qquad (10)$$

where $c$ is the light velocity in vacuum, $v_{gr} = c \bigg/ \left(\varepsilon(\omega_0) + \omega_0 \frac{\partial [\varepsilon(\omega)]}{\partial \omega}\bigg|_{\omega=\omega_0}\right)$ is the group velocity. At a first look it is seems that the flow of energy associated with the first term of the right-hand side of equation (10) (depending on the amplitude $\vec{A}(x,y,z,t)$ of the electrical field) propagates with group velocity, while the second, the magnetic one propagates with the phase velocity. As it is shown below after using the first Maxwell equation, the second term from the right-hand side associated with the magnetic field gives the same amount of energy flow as the first one and propagates with group velocity too.

The next step is to use the differences between the atom and optical scales. The atoms and molecules can be characterized by their atom (molecular) response of order of $\tau_0 \approx 2-3$ fs. During this time the laser pulse propagates at a distance of $z_{resp} = \tau_0 v_{gr} \approx 0.5-1.$ μm. The optical scale is characterized by diffraction length and for a typical laser pulse varies $z_{diff} = k_0 d_0^2 \approx 15-150$ cm. Since $z_{diff} \gg z_{resp}$, always at one diffraction length there are thousand oscillations of the atom dipole. Thus, the shape of the pulse does not change significantly at distances less than one diffraction length and for these distances we can use the following approximation of the Pointing vector:

$$\vec{S} = [0, 0, S_z], \qquad (11)$$

where $z$ is the direction of pulse propagation. In the paper we investigate propagation of optical pulses with linear polarization. In this case the vector amplitudes are orthogonal of the direction of propagation from the conditions $\vec{k}_0 \times \vec{A} = 0$ and $\vec{k}_0 \times \vec{C} = 0$, where $\vec{k}_0 = (0, 0, k_0)$ is the carrying wave-vector and the vector amplitudes are:

$$\vec{A} = [A_x, 0, 0] \quad \vec{C} = [0, C_y, 0]. \qquad (12)$$

Equation (10) transforms to:

$$\frac{\partial S_z}{\partial z} = -\frac{1}{4v_{gr}}\frac{\partial |A_x(r,t)|^2}{\partial t} - \frac{\mu(\omega_0)}{4c}\frac{\partial |C_y(r,t)|^2}{\partial t}. \qquad (13)$$

The equation (13) presents actually the flow of energy through a plane surface situated at point $z = 0$ and orthogonal to the direction of pulse propagation. The coordinates of the intensity of the pulse form the left hand side of this surface are $z - v_{gr}t; \quad t > 0$, while from the right-hand side are $z + v_{gr}t$; $t > 0$. After integrating equation (13) from the left side of this plane and using the fact that the result from the right side is the same, we obtain the following expression of the Pointing vector:

$$S_z = -\frac{1}{2v_{gr}} \int_{-\infty}^{0} \frac{\partial |A_x(x,y,z-v_{gr}t)|^2}{\partial t} dz - \frac{\mu}{2c} \int_{-\infty}^{0} \frac{\partial |C_y(x,y,(z-v_{gr}t))|^2}{\partial t} dz \qquad (14)$$

Thus, from equations (3) and (14) the longitudinal part of the ponder-motor force density connected with the Pointing vector becomes $F_z = {4\pi\chi^{(1)}}/{c}\, \partial S_z/\partial t$ or

$$F_z = -\frac{2\pi\chi^{(1)}}{cv_{gr}} \int_{-\infty}^{0} \frac{\partial^2 |A_x(x,y,z-v_{gr}t)|^2}{\partial t^2} dz - \frac{2\pi\chi^{(1)}\mu}{c^2} \int_{-\infty}^{0} \frac{\partial^2 |C_y(x,y,z-v_{gr}t)|^2}{\partial t^2} dz. \qquad (15)$$

In equation (15) the longitudinal part of the force density is proportional to the second derivatives of the electrical and magnetic pulse envelopes. In cw regime these derivatives vanished and there are only transverse gradient forces investigated in [1, 2]. In femtosecond region, as we will see below, this force takes significant values.

### Longitudinal radiation force density and potential in approximation of first order dispersion

The difference between the atomic and optical scales gives us the chance to solve the integral of force density (15), because at few centimeters the shape of the pulse is practically preserved. The solution of initial Gaussian pulse in approximation of first order dispersion in the frame of spatio-temporal paraxial optics, at distances smaller than diffraction and dispersion lengths is:

$$A_x(t,x,y,z) = A_0 \exp\left(-\frac{x^2+y^2}{2d_0^2} - \frac{(z-v_{gr}t)^2}{2z_0^2}\right), \qquad (16)$$

where $d_0$ is the pulse spot and $z_0 = v_{gr}t_0$ is it's longitudinal shape. To obtain the influence of the magnetic field on the longitudinal force density we use the first Maxwell equation [10]:

$$\nabla \times \vec{E} = -\frac{\mu}{c}\frac{\partial \vec{H}}{\partial t}. \qquad (17)$$

In paraxial approximation the $z$ component of the magnetic field vanishes and equation (17) is simply:

$$\frac{\partial H_y(x,y,x,t)}{\partial t} = -\frac{c}{\mu}\frac{\partial E_x}{\partial z}. \qquad (18)$$

Having in mind that the solution of the electrical field is $E_x = A_x(t,x,y,z)\exp\left[ik_0(z-v_{ph}t)\right]$, where $v_{ph} = c/n_0$ is the phase velocity, $n_0 = n(\omega_0) = \sqrt{\varepsilon\mu}$ is the refractive index and $\varepsilon$ is the dielectric constant, from equation (16) and equation (18) we obtain exact solution of the magnetic field:

$$H_y = H_0 \exp\left[-\frac{x^2+y^2}{2d_0^2} - \frac{v_{ph}}{v_{gr}}\frac{(z-v_{gr}t)^2}{2z_0^2}\right]\exp\left[ik_0(z-v_{ph}t)\right], \qquad (19)$$

where $H_0 = C_0 = A_0 c/\mu v_{ph}$. As it can be seen from solution (19), the square of amplitude of the magnetic field again is Gaussian but slightly deformed with factor $z_0^{mac} = z_0\sqrt{v_{gr}/v_{ph}}$ or:

$$|C_y(t,x,y,z)|^2 = C_0^2 \exp\left(-\frac{x^2+y^2}{d_0^2} - \frac{(z-v_{gr}t)^2}{z_0^2}\frac{v_{ph}}{v_{gr}}\right). \qquad (20)$$

After substituting the square of the amplitudes from equations (16) and (20) in equation (15), differentiating twice by time and integrating by the variable $z$ we obtain:

$$F_z = -\frac{2\pi\chi^{(1)}}{c}A_0^2 \exp\left(-\frac{x^2+y^2}{d_0^2}\right)\frac{t}{t_0^2}\left[\exp\left(-\frac{t^2}{t_0^2}\right) - \frac{2\pi\chi^{(1)}}{c}\frac{\varepsilon}{n}\exp\left(-\frac{t^2}{t_0^2}\frac{v_{ph}}{v_{gr}}\right)\right]. \qquad (21)$$

The paraxial optics works in spatio-temporal coordinates. That is why the expression for PM density force is in the same (x, y, t) coordinates. To present PM force in Cartesian (x, y, z) coordinates we use the relations $z = v_{gr}t$ and $z_0 = v_{gr}t_0$. In this way we obtain the real 3D shape of the radiation force. In addition we present the squared modulus of electrical field by the intensity $|A_0|^2 = 2\pi I_0/cn(\omega_0)$. The expression for the PM density force in a real 3D space is transformed to:

$$F_z = -\frac{4\pi^2\chi^{(1)}v_{gr}}{c^2 n^2}I_0 \exp\left(-\frac{x^2+y^2}{d_0^2}\right)\frac{z}{z_0^2}\left[n\exp\left(-\frac{z^2}{z_0^2}\right) + \varepsilon\exp\left(-\frac{z^2}{z_0^2}\frac{v_{ph}}{v_{gr}}\right)\right]. \qquad (22)$$

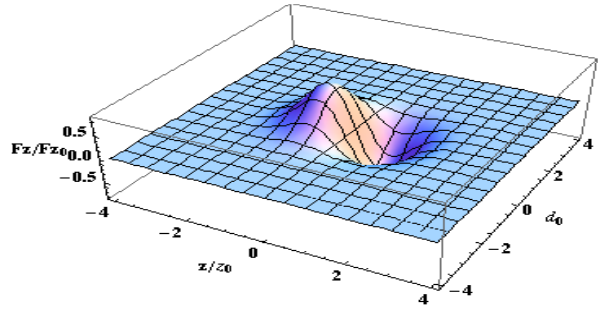

Fig. 1. Graphics of the PM longitudinal force density of a laser pulse. The pulse front attracts the ensemble of particles to the center of the pulse while the back side pushes them again to the center.

The longitudinal PM force propagates with group velocity. The 3D image of the PM force density is plotted in Fig. 1. The pulse front attracts the ensemble of particles to the center of the pulse while the back side pushes them again to the center. The $F_z$ force depends on 3D coordinates and a potential density can be introduced naturally by:

$$U(x,y,z) = \int_{-\infty}^{z} F_z dz .\tag{23}$$

The result is:

$$U_z = -\frac{2\pi^2 \chi^{(1)} v_{gr} I_0}{n^2 c^2} \exp\left(-\frac{x^2+y^2}{d_0^2}\right)\left[n\exp\left(-\frac{z^2}{z_0^2}\right) + \varepsilon \frac{v_{ph}}{v_{gr}} \exp\left(-\frac{z^2}{z_0^2}\frac{v_{ph}}{v_{gr}}\right)\right] .\tag{24}$$

Graph of the potential density is plotted in Fig 2. The Gaussian shape of the pulse plays the role of an attractive potential.

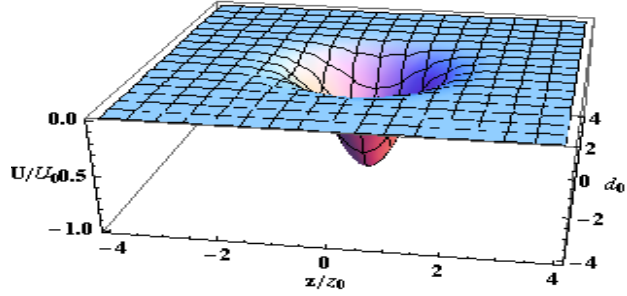

Fig. 2. Graphics of the potential density of a Gaussian laser pulse. The shape of the pulse, moving with the group velocity and plays the role of an attractive potential.

### Values of the longitudinal radiation force and potential

In previous section we obtained formulas for PM longitudinal radiation force and potential densities in approximation of first order of dispersion of a Gaussian laser pulse. To obtain real measurable forces from the density ones, the formulas for the force density (22) and the potential density (24) must be integrated over the whole space. As a result, after integration, we will obtain two additional constants - the spot of the pulse $d_0$ and the longitudinal shape $z_0 = v_{gr} t_0$. In this way an effective real force and potential at level of the spot diameter $d_0$ of the pulse and in the frame of its longitudinal shape $z_0$ is obtained.

After integrating (22) over space for the longitudinal force we have:

$$F_z^{eff} = \iiint\limits_{\substack{x,y=-\infty;\infty \\ z=0;\infty}} F_z dxdydz$$

$$F_z^{eff} = F_z^E + F_z^M = -\frac{2\pi^3 \chi^{(1)} v_{gr}}{nc^2}\frac{E_0^{laser}}{t_0} - \frac{2\pi^3 \chi^{(1)} v_{gr} \varepsilon}{n^2 c^2}\frac{v_{ph}}{v_{gr}}\frac{E_0^{laser}}{t_0} = -Const\frac{E_0^{laser}}{t_0} ,\tag{25}$$

where $F_z^E$ and $F_z^M$ are the electrical and magnetic parts of the averaged PM force, while $E_0^{laser}$ is the energy of the initial laser pulse. It is important to mention here that the longitudinal PM force is proportional to the initial energy and inversely proportional to the pulse time duration. The expression of the potential after integration becomes:

$$U_z^{eff} = F_z^E z_0 + F_z^M z_0^m = -\frac{\pi^3}{2} \frac{\chi^{(1)}}{n(\omega_0)} \frac{v_{gr}^2}{c^2} \left[1 + \left(\frac{v_{ph}}{v_{gr}}\right)^{\frac{3}{2}}\right] E_0^{laser}. \qquad (26)$$

How deep is the radiation potential in air for example? Let's compare it to the Boltzmann energy of free particles at room temperature $T = 300\ K$. The expressions below are written in MKS units ($\chi_{MKS}^{(1)} = 4\pi \chi_{gaussian}^{(1)}$). The value is:

$$U_B = k_B T = 4.14 \times 10^{-21}\ [J]. \qquad (27)$$

In our example we use laser pulse having initial energy in the range of $E_0^{laser} \cong 1\mu J$. The potential is:

$$U_z^{eff} \cong 2.7 \times 10^{-8}\ [J], \qquad (28)$$

which is thirteen orders of magnitude greater than the Boltzmann energy. The Boltzmann factor is very small:

$$R = \exp\left(-\frac{U_{max}}{k_B T}\right) << 1. \qquad (29)$$

This results show that self-confinement of particles into the pulse envelope is possible. Let us suppose that in gaseous media the particles are really confined into the pulse envelope. Then interesting linear and nonlinear effects can be observed. The dipole interaction of the moving neutral particles with the electromagnetic field will be at the carrier to envelope frequency $\omega_{CEF} = k_0(v_{ph} - v_{gr})$ instead at the main ones $\omega_0 = k_0 v_{ph}$. This oscillation is in sub-THz range in gases and can be measured in a direction orthogonal to the direction of the laser pulse propagation. The dipole oscillation measured in the direction of propagation will be again with carrying frequency $\omega_0$ due to the Doppler effect. In nonlinear regime the neutral moving particles will not generate at third harmonics $3\omega_0 = 3k_0 v_{ph}$ but at frequency proportional to the three times group-phase velocity difference $3\omega_{THz} = 3k_0(v_{ph} - v_{gr})$. The influence of moving particles on the four wave-mixing conjugation process was also established in [11].

How strong is the force in fused silica for example? The typical molecular forces in silica are of order of :

$$F_{mol}^{silica} \approx -10\ \left[\frac{eV}{A°}\right] \cong -1.6 \times 10^{-8}\ [N]. \qquad (30)$$

If we use a laser pulse having time duration $t_0 = 100$ $[fs]$ and energy $E_0^{laser} \cong 100$ $nJ$ the value of the longitudinal force becomes:

$$F_{pm}^{pulse} \cong -7.02 \times 10^{-2} \quad [N]. \tag{31}$$

This value is six's order of magnitude greater than the molecular forces in silica. This calculations show that the fine ablation obtained by femtosecond pulses in silica can be result of broken molecular connections due to longitudinal radiation force.

### Discussions

Up to now, the basic experimental and theoretical investigations are related to the study of radiation forces produced by laser beams and pulses acting on individual Rayleigh dielectric particles. In this paper we explore the impact of the longitudinal force, associate with the Pointing vector and its influence on an ensemble of particles in dielectrics. Thus, the individual force applied to an atom is transformed to density force per volume. The optical response of dielectric media connected with the propagation of laser pulses is non-stationary and also is taken into account. As a result, analytical expression for the longitudinal force density and potential density of Gaussian pulse are obtained. It is possible to integrate these densities using the paraxial approximation in optics. Measured in the experiments average effective longitudinal potential and force, acting with particles at the level of the pulse spot is presented for a first time. The longitudinal radiation force is proportional to the initial pulse energy and inversely proportional to the pulse time duration. In the femtosecond region this force in silica for example is of few orders of magnitude greater than the molecular forces. Therefore, the fine ablation with fs pulses in silica can be realized by broken molecular connections due to this longitudinal PM force. The force is of potential type and in air, for an example, the potential of a Gaussian laser pulse with energy $E_0^{pulse} \cong 1.$ µJ is thirteen orders of magnitude greater than the Boltzmann energy of free particles. It is possible that the neutral particles to be confined in the pulse envelope and to move with group velocity. Then, the dipole interaction of the moving neutral particles with the electromagnetic field will generate wave at the carrier-to-envelope frequency instead at the main ones. This oscillation is in sub-THz range in gases and can be measured in a direction orthogonal to the direction of propagation of the laser pulse. The measured dipole oscillation in direction of propagation will have again the carrying frequency due to the Doppler effect. In nonlinear regime the neutral moving particles will not generate at third harmonics but at frequency proportional to the three times group-phase velocity difference. Such generation was indeed observed in recent experiments [12].

#### Funding
This material is based upon work supported by the Air Force Office of Scientific Research under award number FA9550-19-1-7003. The present work is funded also by Bulgarian National Science Fund by grant DN18/11.

#### Disclosures

The authors declare no conflicts of interest.